# Urban Data Streams and Machine Learning: A Case of Swiss Real Estate Market

Vahid Moosavi
Chair for Computer Aided Architectural Design, Department of Architecture
ETH Zurich
Zurich, Switzerland
svm@arch.ethz.ch
www.vahidmoosavi.com

*Abstract*—In this paper, we show how using publicly available data streams and machine learning algorithms one can develop practical data driven services with no input from domain experts as a form of prior knowledge. We report the initial steps toward development of a real estate portal in Switzerland. Based on continuous web crawling of publicly available real estate advertisements and using building data from Open Street Map, we developed a system, where we roughly estimate the rental and sale price indexes of 1.7 million buildings across the country. In addition to these rough estimates, we developed a web based API for accurate automated valuation of rental prices of individual properties and spatial sensitivity analysis of rental market. We tested several established function approximation methods against the test data to check the quality of the rental price estimations and based on our experiments, Random Forest gives very reasonable results with the median absolute relative error of 6.57 percent, which is comparable with the state of the art in the industry. We argue that while recently there have been successful cases of real estate portals, which are based on Big Data, majority of the existing solutions are expensive, limited to certain users and mostly with non-transparent underlying systems. As an alternative we discuss, how using the crawled data sets and other open data sets provided from different institutes it is easily possible to develop data driven services for spatial and temporal sensitivity analysis in the real estate market to be used for different stakeholders. We believe that this kind of digital literacy can disrupt many other existing business concepts across many domains.

I. INTRODUCTION

The traditional paradigm of data management and business analytics in the real estate market has been that usually in each region or country there are few dominant players with access to a restricted source of data that are providing relatively expensive and limited business analytics solutions. Historically, this is based on the fact that the real estate data including transaction data has never been publicly available.

As a result of this closed ecosystem of data management and business analytics, usually decision-making is happening either in an isolated environment or based on non-transparent indexical systems imposed by consulting companies and usually used for example in banks. Therefore, while individuals (e.g. sellers, buyers, tenants and landlords) are metaphorically producing the drops of water in the stream of real estate data, they usually do not gain from the out coming Big Data.

Based on author's opinion, the current stage in real estate market is very similar to the era before the democratization of geo-mapping, specially before the introduction of Google Earth and Google Map services, where the notion of spatial information, was something associated with the domain experts who know how to work with maps and geo-data. As a result, due to lack of communication among different kinds of stakeholders the final geo-products were based on static data and in a lower quality compared to current geo-products. This can be exemplified by the case of traffic information on Google maps, which is nowadays being updated continuously from the aggregated online trails of individuals who are moving across the road networks, while the GPS of their mobile devices are switched on and are sharing their data. In return, Google map gets updated more realistically and eventually every body will get a better routing service based on a dynamic road map in comparison to classically static maps. In the real estate market, similar set ups could open up new dynamics. For example, in the case of automated property evaluation if there is a dynamic game-like environment, where the system initially suggests an estimated price index for a region or a single property and then the stakeholders (e.g. investors, agents, potential tenants, landlords, etc.) are continuously giving negative or positive feedbacks to the initial estimates, not only can the system become more accurate and data driven, but also, gradually over time the interactions between users and the system will lead to a new source of data which in principle, indicates the willingness to pay and demand dynamics in comparison to the supply dynamics. This has been the conceptual idea behind this work and at the moment a new web portal with this principle is under development for the Swiss real estate market, called FairyFy.com[1].

Toward this goal, in this paper we report the initial steps in developing such a system. We start by describing the data collection pipeline, where by crawling publicly available real estate advertisements from an existing real estate portals in Switzerland we dynamically collect several types of advertised items for rental or sales across the country. Further, considering the crawled listings as spatiotemporal samples, we frequently provide rough estimates of the price index of all the buildings in Switzerland, using the publicly available data from Open Street Map (OSM) and other publically available data sources. Next, using different well-established machine learning and function approximation techniques we show how easily we developed a nonlinear data driven method for individual property evaluations. Further, we show how we develop different data driven analytics tools, where the users can interactively perform spatial sensitivity analysis of the real estate market at the zip code level. Finally, we discuss the main challenges and opportunities by sketching the conceptual framework of the proposed real estate portal.

---
[1] The working prototype of this website is currently accessible at http://www.fairyfy.com/

## II. Data Collection Pipeline

In this work, we used two main data sources. First, in order to get the information about residential buildings, we simply extracted the building layer of OSM data for Switzerland. Secondly, since usually real estate data such as transactions or advertised listings are not easily accessible as free Big Data collections, we wrote a web crawler using http requests and XML parsers that runs continuously on a mini-PC and on a daily basis collects all the publically available advertisements in one of the main real estate portals of Switzerland, called Homegate. There are different categories of listing, for both rental and sales items that need separate http requests. Further, we should note that since the procedure for rental and sale listings are all the same, in this work we only focus on the rental data sets.

The crawled data has several attributes including a unique ID, zip code, address, type of property, number of rooms, floor level, size of living space, size of floor space, room height, volume, year built, last renovation, net rent, additional expenses, rent, date of availability, view (e.g. lake view), balcony/patio, fireplace, cable TV, ISDN connection, children welcome, parking place, Garage, wheelchair accessibility, allowance of pets, elevator, group living arrangement, new building, old building, swimming pool, location, public transport distance, shopping distance, kindergarten distance, primary school distance, secondary school distance, motorway distance, floor plan document and verbal descriptions.

However, for each listed item, usually there are several missing elements, for which we developed a non-parametric probabilistic imputation method that its description goes beyond the scope of this paper. Nevertheless, to test the quality of the automated property valuation systems, we only use those items with complete fields. After data cleaning, we perform the geocoding step, using Google geocoding API, where by sending the address of a listed item to the server, we get its geo-coordinates.

On a daily basis, there are around 1500-1600 new rental items, posted to the website, from which on average 1200-1400 are new residential items. The current strategy of the crawled real estate portal is to remove the records of old advertisements. Therefore, we need to crawl this data frequently. Accumulation of these data streams over time will be a valuable data source that captures the dynamics of real estate supply side. In the next sections, we describe the main applications we developed based on these crawled data sets. Further, we should note that since the collected data sets are based on publically available data, we frequently upload the cleaned data sets into our website, where the visitors can easily download both rental and sales items for free (shown on top of Fig. 1 in a separated tab). Further, as shown in Fig. 1 the original advertisements are plotted into the developed geo-map, where by clicking on them the user will be transferred to the original advertisement. However, based on our experiences using geo-visualizations, it is easily possible for users to compare the conditions of one specific property in comparison to other listings in its vicinity, which gives a better overview of the neighborhood to the person searching for a property.

## III. Rough Estimations of the Price Index of 1.7 Million Buildings in Switzerland

As mentioned before, this paper is partially reporting the development of a free public website, where users can use it like the original real estate portal, where the data is crawled rom. In addition, in our developed website, which is currently online there will be several new data driven features that attract the public attention to the system. In this regard, one of the first ideas was to estimate the rental price of all the residential buildings in Switzerland. Toward this goal we started working with OSM data source, which is freely accessible.

OSM layers have in principle extremely rich information about the objects in each layer. For example, at the building layer, each building has different features such as address, geo-location, name, land use, age, 3D geometric information and as a result the building footprint, building size and building height. Further, using information of other layers such as points of interests, places and road networks, it is nowadays possible to estimate several types of environmental factors for each property such as the distance of each building to the nearest public transport station, that are usually important in order to assess the quality of real estate object.

However after extracting the building layers of Switzerland, which consists of around 1.7 million buildings, we realized that the building information in Switzerland is not yet that complete in comparison. For example, building footprints are not always correct and in many cases a big polygon, which consists of several buildings is considered one building. As a result, instead of predicting the exact rental price of a building, which would be based on its estimated area and location, we decided to predict the rental price per square meter for each of individual buildings using only the geo coordinates of the buildings. Considering the crawled data set as a kind of spatial samples, we fitted a nonlinear local function to estimate the rental price index of each building. Similar to [7] we trained a Self Organizing Maps (SOM) [8] based on only latitude, longitude and price per sq. meter of all the crawled advertisements. The trained SOM finds nonlinear clusters of locations, which have the same price patterns. Specifically, in this case, SOM learns a non-parametric probability distribution over geo coordinates and possible prices. For other buildings with only geo coordinates, we sample over the trained SOM, using K-

Nearest Neighborhood (KNN) method. In order to choose a price value for a specific data point, in principle, there will be two strategies. The first one, which is common, is to select the median price values of the trained nodes in the K selected nodes. The second one is to draw samples over those training data points, which belongs to the K selected nodes of SOM. While, the second approach would better conserves the probability distribution of the training data, the first approach produces more smooth values that visually look better. Therefore, in this work we chose the second approach, as our goal was not to come up with an exact estimate for each building, but to provide coherent and smooth local price estimates.

We visually compared the quality of predictions with the color-coded values of spatially clustered advertised items. We checked these results with some domain experts who know the real estate market of Switzerland very well, where they found these estimates as visually convincing. Fig. 1 shows a snapshot of the developed web page, visualizing the estimated rental index for the region around city of Geneva. Colored circles are corresponding to a spatial cluster of listed rental items and the shown numbers in each circle correspond to the aggregated rental price index of that cluster.

## IV. ACCURATE ESTIMATION OF THE RENTAL PRICE OF A SINGLE PROPERTY

While in the previous section, we showed how rough estimations of price indexes of buildings based only on their geo coordinates gave us visually convincing results, it is obvious that for an accurate prediction of price of a property, more factors and more complex methods will be needed. The problem of single property valuation falls in the class of spatial regression, where in addition to specific features of the item (K.K.A intrinsic features), there are environmental factors that contribute to the values of the predicted features. Therefore, while the specific features of a house mainly determine the price of a house, there are always some sorts of correlations between the prices of neighboring houses.

In the context of economic and real estate models, the problem of valuation has a long history. For example one can refer to [5] and [6]. Among very common economic approaches one can refer to a class of linear regression models, which are called hedonic regression [8], where the goal is to decompose the price of a house to several contributing factors or hedonic indexes. For example, ideally the price of a house is a weighted sum of its hedonic indexes such as the number of rooms, living space, floor number, building age, etc. In this way, the hedonic models give lots of freedom to the economists to perform different kinds of policy analysis.

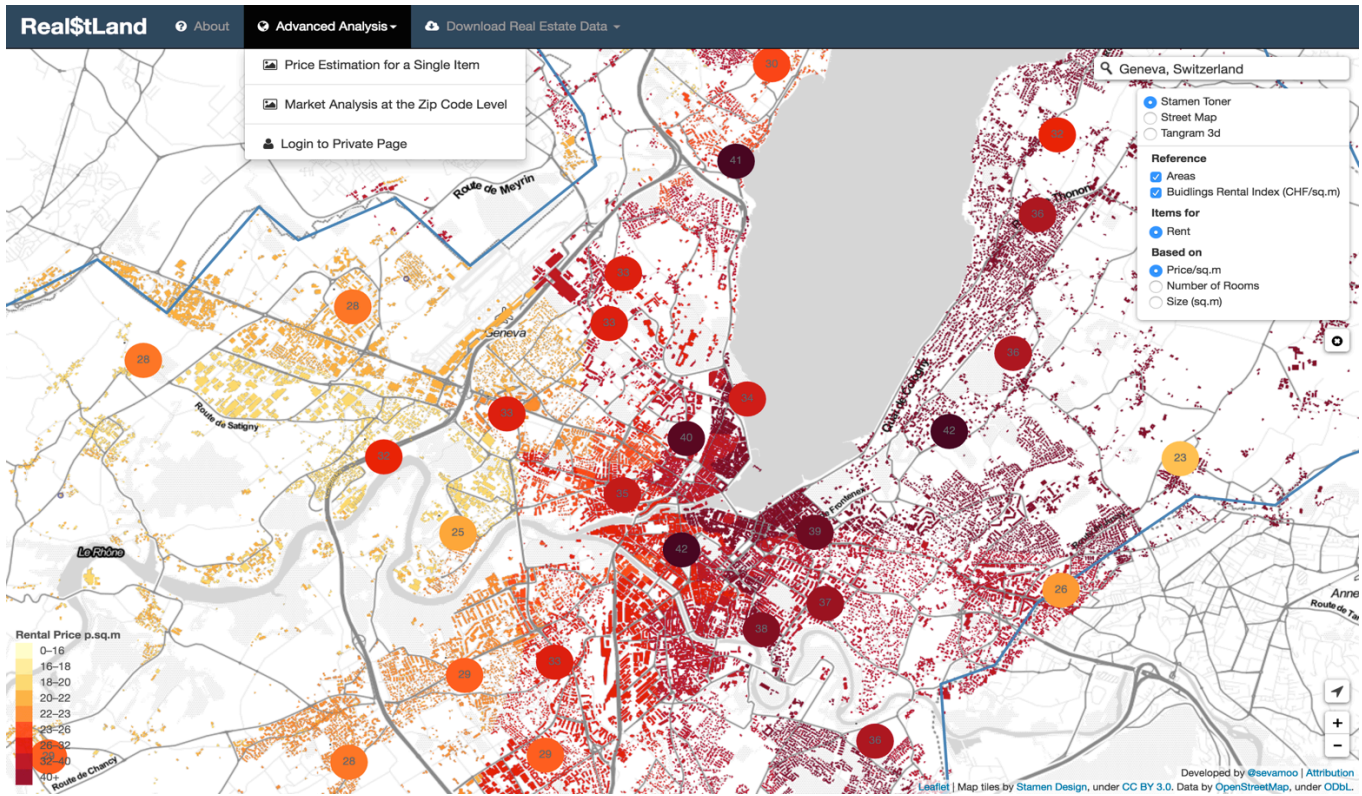

Figure 1. A snapshot of the developed web site, where buildings in Geneva, Switzerland are colored based on their estimated rental price per square meters and the colored circles are local clusters of the listed advertisements. By clicking on each cluster, the user can see the individual advertisements, where each individual advertisement has a link to its original page in the crawled real estate portal.

The literature of hedonic models is very extensive. For example, there have been some modifications to the classical linear models by introduction of nonlinear functional forms, which are based on logarithmic transformations of original observed values. Among these transformed hedonic models, Box-Cox models [1] offer the highest level of flexibility and are the most commonly used methods in the literature.

However, as it is discussed in [3] if we look at the problem of property valuation as a function approximation problem, with not necessarily decomposable structures, there are many machine learning methods available that can outperform the current linear models in terms of predictability. This is opposite to the majority of current economics literature, where the goal is to construct an efficient and structured index system that is used for policy analysis. Therefore, since in this work our goal is to develop an automatic evaluation system that estimates the price index of any house in response to specific user queries, our focus should be on developing a reliable function approximation system rather than focusing on the structure of the price in real estate market. In this work, we performed several experiments including K-nearest Neighborhood (KNN) and Random Forest method [2] and different types of linear regressions.

We should note that hedonic linear regression methods such as Box-Cox were not tested here, since based on the experiments in [3], the result from KNN was significantly superior to the results from Box-Cox method. Nevertheless, in order to have an error base line, we tested the linear regression; Bayesian regularized regression [9] and the weighted local polynomial regression with different polynomial orders.

As we explained before, the crawled data set has several features that both describe the property itself as well its environmental factors. However, as majority of the advertised items do not have all the environmental information, there was a tradeoff between having more features with less training data sets and having more training data sets with less features. We tested all the methods with different feature sets, but finally we concluded to keep only the following features: Type of the property, number of rooms, floor level, size of living space, year built, zip code, longitude and latitude of the building. We should note that we converted the type of the property from a categorical variable with four values of Apartment, Duplex, Single house and Studio to one-hot 4 dimensional vectors.

Further, in order to visually explore the relationships between possible features and the rental prices, we trained several SOM using the training data with the selected features. For example, Fig. 2 shows that there are several non linear, but local relationships between the selected features. The colors show the normalized values within each feature. Therefore, if there are some local or global interrelations it can be quickly visualized on the trained SOM. For example, as it was logically expected, the number of rooms and living space are correlating globally, while there is a specific cluster in the middle of the maps, where the rent is high and it correlates with a certain range of room numbers, living space and a specific location in Switzerland, which is highlighted in the corresponding maps for zip code, longitude (lng) and latitude (lat). However, in principle by an appropriate selection of a nonlinear function approximation method, we expect to grasp all these nonlinearities.

In this paper, we only chose the data crawled in June 2016, which consists of 23,854 unique rental properties, after removing the rows with at least one missing values.

In order to test the quality of predictions, we divided the data randomly to 80% training data and 20% test data and we performed out of sample predictions for several times. The quality of the predictions was measured in terms of the Absolute Relative Error (ARE) [4]. Let $A$ be the actual rental price of a house, and let P be its predicted price. Then the Absolute Relative Error is defined as

$$ARE = |A-P|/A \qquad (1)$$

We implemented all the function approximation methods, using the machine-learning package scikit-learn [10], PyQt-Fit for the weighted local regression and SOMPY for SOM.

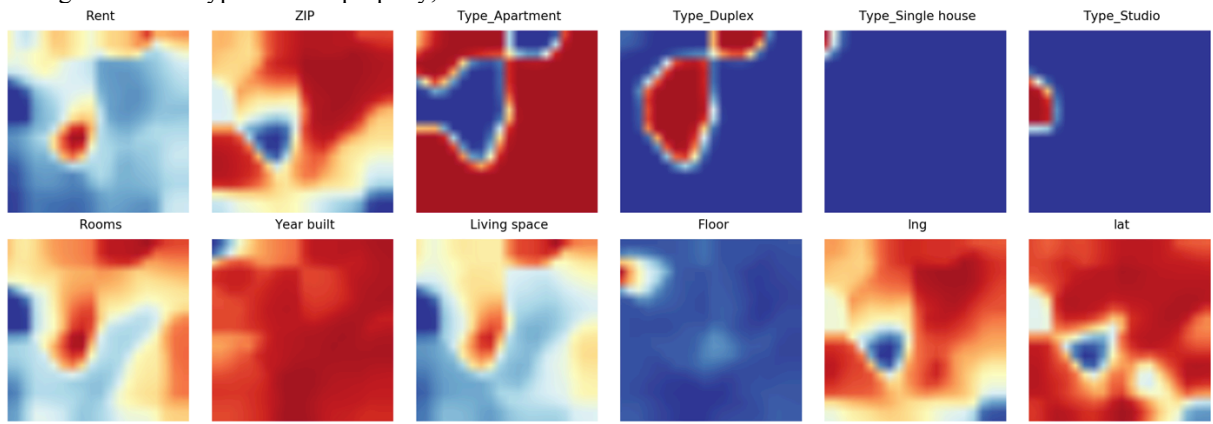

Figure 2. A trained SOM visually unfolds the interrelationships between different features of rental items. The colors show the normalized values within each feature.

In order to find the best meta-parameters, we tested several combinations of meta-parameters for each method. Finally, in Random Forest we chose the number of estimators (i.e. the number of decision trees) to be 80 and the other meta-parameters as the default values of the library.

For KNN with K=9, we got the best results in several runs. For Linear Regression and Bayesian Regularized Regression we used the default parameters of the library that automatically finds the optimum regression and regularization coefficients. For weighted local polynomial regression, we tested several polynomial orders, but In general, higher polynomial orders will over fit to the data. Therefore, we tested only orders 1,2 and 3.

In order to compare the prediction quality of different methods, in addition to median error, three other performance quantities are reported. We show the percentage of houses with the prediction errors of less than several error thresholds including 1%, 5% and 15% in order to show the distribution of errors in addition to the average values, which might be biased by the outliers. Therefore, the greater these values the better the system.

TABLE I. shows these values for different predictors. As we expected, Random Forest and KNN give the best results, while Random Forest is giving slightly better results than KNN. Further, in comparison, local regression with polynomial functions obviously give much better results than the global linear regressions. In addition, it is clear that in the case of local regression methods, polynomial order of 2 gives a better distribution of error than polynomial order of 3, though in terms of median error and errors less than 1% polynomial order 3 gives better results.

TABLE I. PREDICTION ERRORS OF DIFFERENT ALGORITHMS ON TEST SET INCLUDING THE MEDIAN ERROR AND THE PERCENTAGES OF TEST DATA WITH DIFFERENT LEVELS OF ERRORS FROM 1% TO 15%

| Algorithm | Median | <=1% | <5% | <15% |
|---|---|---|---|---|
| **Random Forest** | **6.57** | 9.64 | **40.12** | **78.83** |
| **KNN** | 7.17 | **10.02** | 38.98 | 76.33 |
| **Bayesian Regularized Regression** | 13.89 | 3.74 | 19.86 | 52.76 |
| **Linear Regression** | 13.92 | 3.81 | 19.86 | 52.79 |
| **Local Regression P-Order=1** | 9.34 | 5.74 | 29.40 | 68.71 |
| **Local Regression P-Order=2** | 8.53 | 6.52 | 32.19 | 71.81 |
| **Local Regression P-Order=3** | 8.45 | 6.62 | 31.98 | 70.74 |

The ultimate goal of this step was to deploy the implemented evaluation system in a web based API, where we regularly train and update the best prediction system and using the trained system, the users of the system will make queries about specific houses or their own properties by providing all the required features of a property used in the function approximation system. However, since the system is only based on a limited feature set we would let the users to give feedbacks to the results based on their personal expectations. For example, one possible scenario would be to give several options to the users to select as a reason why the estimated price is higher or lower than their expectations. In this way, we hope that rather than providing a top-down and expert based price estimation, we develop a mechanism for data collection through users queries and later by adding new features, which might be important, and currently missing in the existing data set. In addition, public reaction to the estimations could be collected systematically, which would show the willingness to pay in the market, regardless of the market supply price. This for example, can be of enormous value for those property owners who want to know about the opinions of the visitors about their houses. In long term, this scenario will bring more dynamics to the current real estate market, where there are no direct interactions between supply and demand sides.

## V. SENSITIVITY ANALYSIS OF THE REAL ESTATE MARKET

On a daily basis the current real estate portals in Switzerland receive lots of new listings and usually due to the massive size of these data streams they usually do not keep all of these data in their databases. As a result, the current portals in Switzerland are usually offering a plain temporal list of advertisements in response to queries of people who are looking for a property. In this section, we show how using the collection of these data streams from a certain period of time (e.g. a month or a quarter of a year) we can develop simple interactive analytics that give insights to different stakeholders. In the previous section, we showed how the implemented data driven system was able to give an estimated price for a specific property. However, in an inverse scenario a potential tenant/buyer is usually interested to know where there are higher chances of finding a house, which matches very well with his/her expectations. Or if based on a certain budget there are very low chances in a region, how much one should add to the budget in order to increase the chance of renting a house in that certain area. Or in another scenario, a construction company might need to know about the distribution of rooms, sizes and prices for a certain region or considering the time dimension, one might be interested to know the development of the average price of a region over a certain period of time. Based on these scenarios, it is easily possible to develop different services using only the collected data sets. In this work, we only present two use cases that we developed so far. In the first use case, the user makes a query by selecting a range of values for three main parameters of a house including the number of rooms, the size of living space and the monthly rent, which shows his/her willingness to pay. As a result, the system gives the user a colored map showing where the user has more options and higher chances of finding a place based on the selected query. Then, the user can interactively change the queries to see the sensitivity of different regions to different parameters. Fig. 3 as an example shows the percent of available rental cases in each zip code of Switzerland based on the query of min 3 rooms, min 50 square meters of living space and maximum 3000 CHF per month for the month June 2016. In this use case, the system gives an overall picture of the market to the user.

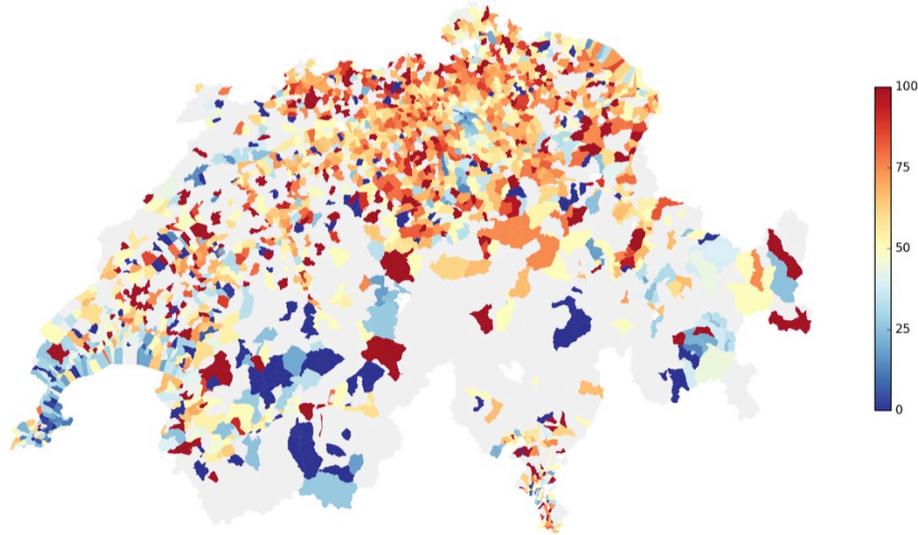

Figure 3. Percent of available rental items in each zip code of Switzerland based on the query of min 3 rooms, min 50 square meters of living space and maximum 3000 CHF monthly rents. The gray areas indicate no data for that region.

In the second use case, we assume that the user is interested in a specific region. Therefore, there will be another option available, where the user can see the distribution of supply in relation to his/her expectations. As another example, Fig. 4 shows the accumulative distribution of matching rental items by increasing the monthly budget for renting a house with at least 2 rooms and minimum of 50 square meters of the living space in 6 zip codes, located in downtown area of Zurich. As it can be seen different zip codes behave differently in response to the change in the monthly rents. In addition to this, Fig. 5 shows the distribution of matching items (i.e. combinations of rooms, size and price) for two different monthly rental budgets of maximum 3000 CHF and 5000 CHF for zip code 8005 in Zurich with the above-mentioned limits for number of rooms and the size of living space. The user can interactively, change these values.

Similar use cases would work for different stakeholders such as landlords or construction companies if they want to optimize the portfolio of their buildings. Further, similar to the interest of economists in hedonic models using these dynamically crawled listings, it is possible to analyze the sensitivity of the rental price of a house in different regions to any parameters of interest such as the number of rooms or the size of living space, floor number, etc. Nevertheless, our goal in this paper was to show the potential applications of these publicly available data. In the developed web based application, we are currently implementing all of these services gradually, which will be freely available for public use.

VI. CONCLUSIONS

This paper is partially reporting the development of a public real estate analytics portal in Switzerland. We showed how only by continuous crawling of publicly available real estate advertisements as a kind of Big Data streams; one could create free valuable services for different stakeholders of real estate market. This approach is conceptually different than the current ecosystem of data management and business analytics in the real estate industry, where while the data is being produced by many individual interactions, the data-driven services are based on expensive and private data collections.

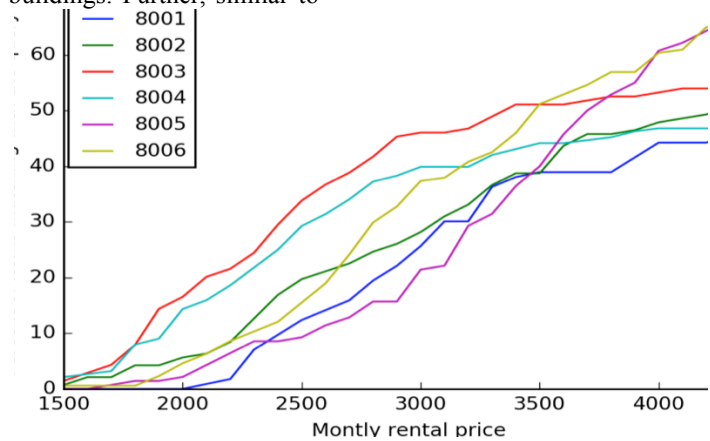

Figure 4. Percent of rental items with at least 2 rooms and 50 square meters of living space that match with different monthly rental budgets for different zip codes in downtown Zurich

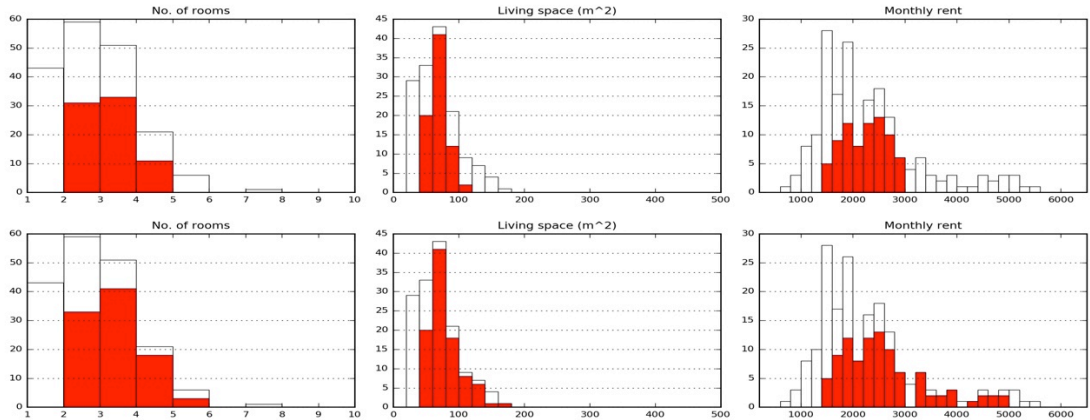

Figure 5. Distribution of available rental cases in zip code 8003 in Zurich, based on two queries of minimum 2 rooms, minimum 50 square meters of living space and the monthly rent of maximum 3000 CHF (top row) and 5000 CHF (button row). White histograms are based the total available cases and red histograms are based on those cases that match with the selected query.

First, by using the crawled data and open source building data from OSM, we showed how to roughly estimate the rental price index of around 1.7 million buildings in Switzerland. Later we discussed how using different function approximation techniques we developed a fully data driven rental price estimation system, which achieves a reasonably low error rate.

Further, we showed that the dynamically crawled data set has other values, where for example one can develop several types of business analytics for different kinds of stakeholders such as potential tenants, landlords, construction companies and economists.

Further, since the whole platform is developed on top of publically available data sources, our ultimate strategy is that unlike current real estate consulting companies we will provide all the services freely to the users of the system. On the other hand, we hope that such frameworks will change the current opaque and closed evaluation mechanisms in the real estate market to a new level that the developed data driven estimates are not developed to dictate the price indices, but rather to create a game like environment, where different stakeholders such as landlords, tenants and investors can react on the estimated values either by criticizing or admitting the predicted prices and by bringing more specific evidence and insights from their individual experiences. Therefore in this sense, an automated valuation system can be seen as an engine for simulating the market dynamics and to collect data from the demand side and willingness to pay, which has a completely different nature than the supply data. Intersection of demand and supply dynamics which is rarely accessible, has a high value for all the stakeholders in this market.